\documentclass[pra, 
floatfix,
twocolumn
]{revtex4}

\usepackage{graphicx}
\usepackage{color}
\usepackage{float}
\usepackage{amsmath, amsfonts, amssymb, bm}
\usepackage{hyperref}
\hypersetup{
	colorlinks=true,
	linkcolor=blue,
	filecolor=magenta,      
	urlcolor=blue,
	citecolor=red
}
\begin{document}
\title{Electron-positron pair production in strong oscillating\\ electric field with multi-pulse structure}

\author{Abhinav Jangir}
\email{2022rpy9087@mnit.ac.in}
\address{Department of Physics, MNIT Jaipur, Jaipur, Rajasthan, India}
\date{\today}
\begin{abstract}
	We investigate electron-positron pair production from the vacuum in presence of a strong oscillating electric field with a multi-pulse structure and variable inter-pulse delay. The pair production probabilities are computed by numerically solving the time-dependent Dirac equation. We analyze the resulting momentum distribution and the total number density of produced particles for different numbers of pulses and inter-pulse delays. In particular, we demonstrate the emergence of a characteristic time-domain multi-slit interference pattern in the pair production probability as a function of the inter-pulse delay.
\end{abstract}

\maketitle

\section{Introduction}\label{sec:intro}	
	In presence of ultra-strong electric fields, the quantum vacuum can become unstable and decay into electron-positron pairs. This remarkable nonperturbative phenomenon, known as the Schwinger effect, stands as one of the profound predictions of quantum electrodynamics (QED). 
	The relativistic theory of the electron in 1928 by Dirac \cite{Dirac1928Quantum} led to the prediction of positron. This paved the way for Sauter \cite{Sauter1931Uber} to demonstrate that a sufficiently strong static electric field can induce tunneling from the negative-energy Dirac sea, resulting in electron-positron pair creation. Later, in 1951, Schwinger gave a complete quantitative description of the vacuum decay and derived the pair production rate in a constant electric field using the proper-time formalism \cite{Schwinger1951On}. His result introduced the critical field strength $E_\text{cr} = m^2c^3/e\hbar \approx 1.3 \times 10^{18}$ V/m, which sets the characteristic scale at which vacuum decay becomes significant. This field strength corresponds to an enormous laser intensity of the order of $10^{29}$ W/cm$^2$, which is still far beyond the reach of present-day laser technologies. Hence, the direct observation of the Schwinger effect a remains a challenge for the theoretical as well as the experimental researchers. The rapid development of next-generation high-intensity laser facilities, such as the Extreme Light Infrastructure \cite{ELI}, Stanford Linear Acceleration Center (SLAC) \cite{SLAC}, and the European X-ray Free-Electron Laser \cite{XFEL}, is expected to bring field strengths closer to the regime required for observing Schwinger-like pair production processes. These advances offer a promising pathway to experimentally probe the nonlinear regime of quantum electrodynamics.
	
	The simplest field configuration used to model laser pulses is a time dependent, periodically oscillating electric field. Such a structure naturally arises in the form of a standing wave, which can be generated through the superposition of two counterpropagating laser pulses. In this configuration, the spatial dependence of the field can be neglected near the antinodes, allowing one to approximate the system as a purely time-dependent electric field. This approximation is widely employed in strong-field QED studies, where it captures the essential features of particle dynamics. 
	The decay of vacuum into the electron-positron pairs in presence of such monofrequent time dependent oscillating electric fields was first investigated in the 1970s \cite{brezin1970pair, Grib1972Particle, Bagrov1975Concerning}. These foundational studies revealed that pair production can occur through distinct interaction regimes, each exhibiting qualitatively different behavior. The classification of these regimes is determined by a dimensionless quantity, the adiabaticity parameter \cite{brezin1970pair}, defined as
	\begin{equation}\label{eq:adiabaticity_parameter}
		\xi = \frac{|e|E_0}{mc\omega},
	\end{equation}
	which is essentially the inverse of the Keldysh parameter \cite{keldysh2024ionization}. Here, $E_0$ is the field strength, $\omega$ is the field frequency, $e$ is the electron charge, $m$ is its mass, and $c$ is the speed of light.
	
	 In the regime $\xi\ll1$, pair production occurs via a perturbative multiphoton process, leading to a power-law dependence of the production probability on the field amplitude $E_0$. In contrast, for $\xi\gg 1$ and subcritical field strengths $(E_0\ll E_\text{cr})$, the process becomes nonperturbative, exhibiting an exponential dependence on the field strength, analogous to the behavior in a constant electric field \cite{Schwinger1951On}. The intermediate regime $\xi\sim1$, represents a transition region where multiphoton and nonperturbative effects coexist. This domain is of particular interest, but also poses significant theoretical challenges, as no simple asymptotic descriptions are available \cite{avetissian2002electron}.
	 
	 Recently, a substantial number of studies have investigated pair creation by considering electric fields of finite duration \cite{piazza2004pair, hebenstreit2009momentum, mocken2010nonperturbative} and different envelope shapes \cite{kohlfurst2013optimizing, aleksandrov2017pulse, abdukerim2013effects, linder2015pulse, blinne2014pair, li2017momentum, unger2019infinite}. Since the required field strengths remain unattainable with current technologies, various methods have been developed to enhance pair production, such as the inclusion of frequency chirps \cite{dumlu2010schwinger, chen2024asymmetric, gong2020electron, olugh2025frequency} and the superposition of a weak, rapidly varying field on a strong, slowly varying field, known as the Dynamically Assisted Schwinger Effect \cite{orthaber2011momentum, linder2015pulse, nuriman2012enhanced, schneider2016dynamically, sitiwaldi2018pair}. The effects of spatial inhomogeneities on pair production have also been explored in addition with other enhancement methods \cite{xu2025effect, schneider2016dynamically}. These studies have shown that pair production is highly sensitive to the precise form of the electric field.
	 
	In pulse-train electric fields, coherent enhancements arising from multiple-slit–like interference in the time domain have been extensively reported \cite{akkermans2012ramsey, li2014enhanced, kaminski2018diffraction, li2014multiple}. These interference effects rise as characteristic oscillatory structures in the pair production probability, reflecting the phase coherence between successive pulses. 
	In particular, Ref. \cite{akkermans2012ramsey} showed that for a sequence of $N$ Sauter-type pulses with alternating-sign, the interference occurs with the central value scaling as $N^2$ relative to the maximum distribution produced by a single pulse. This was compared with the analytical and numerical predictions based on WKB methods and turning-point analysis \cite{dumlu2011interference}.
	
	In this work, we investigate electron–positron pair production in strong, time dependent oscillating electric fields with a multi-pulse structure in the nonperturbative regime of $\xi \sim 1$. Our study extends the analysis of Ref. \cite{granz2019electron}, where only a double-pulse structure was considered. By generalizing to multiple pulses, we examine how the inter-pulse delay and the number of pulses jointly influence the production yield. In particular, we demonstrate the emergence of a characteristic time-domain multi-slit interference pattern in the pair production probability as a function of the inter-pulse delay. The corresponding time-dependent Dirac equation is solved numerically to compute both the momentum-resolved production probabilities and the total number of produced pairs.
	
	The rest of this paper is organized as follows. In Sec.~\ref{sec:computational_framework}, we introduce the electric field model and outline the numerical framework based on solving the time-dependent Dirac equation. In Sec.~\ref{sec:numerical_results}, we present and discuss our numerical results. Finally, we summarize our findings in Sec.~\ref{sec:conclusion}. Throughout this work, relativistic units with $\hbar = c = 1$ are used. 

\section{Computational framework}\label{sec:computational_framework}
	\subsection{Field profile}\label{subsec:field_profile}
	Our goal is to study electron-positron pair creation in a time-dependent, oscillating electric field with a multi-pulse structure. We consider a sequence of $K$ linearly polarized laser pulses along the $y$-direction, each having a smooth envelope and a sinusoidal carrier. Every pulse contains $N$ cycles of frequency $\omega$, and successive pulses are separated by a time delay $\delta$. The construction is performed in the temporal gauge, so that the electric field $\mathbf{E}(t) = -\dot{\mathbf{A}}$ can be expressed in terms of the vector potential as
	
	\begin{equation}
		\mathbf{A}(t) = \sum_{k=0}^{K-1} \frac{m \, \xi}{e} \, \sin\!\big[\omega (t - t_k)\big] \, F(t - t_k; \tau_p, \tau)\hat{y},
		\label{eq:A(t)_vector_potential}
	\end{equation}
	
	where $t_k = k (\tau_p + \delta)$ is the start time of the $k$-th pulse, $\xi$ the dimensionless intensity parameter, the pulse duration $\tau_p =  (N+1)\frac{2 \pi }{\omega}$ and the switching time $\tau=\frac{\pi}{\omega}$.
	
	The envelope function $F(t; \tau_p, \tau)$ has compact support on $[0,\tau_p]$ and smoothly modulates the pulse amplitude:
	
	\begin{equation}\label{eq:envelope}
		F(t; \tau_p, \tau) =
		\begin{cases}
			\sin^2 \left(\frac{\omega t}{2}\right), & 0 \le t < \tau, \\
			1, & \tau \le t < \tau_p - \tau, \\
			\sin^2 \left(\frac{\omega (\tau_p - t)}{2}\right), & \tau_p - \tau \le t \le \tau_p, \\
			0, & \text{otherwise}.
		\end{cases}
	\end{equation}
	
\begin{figure}[tbh]
		\includegraphics[width=0.5\textwidth]{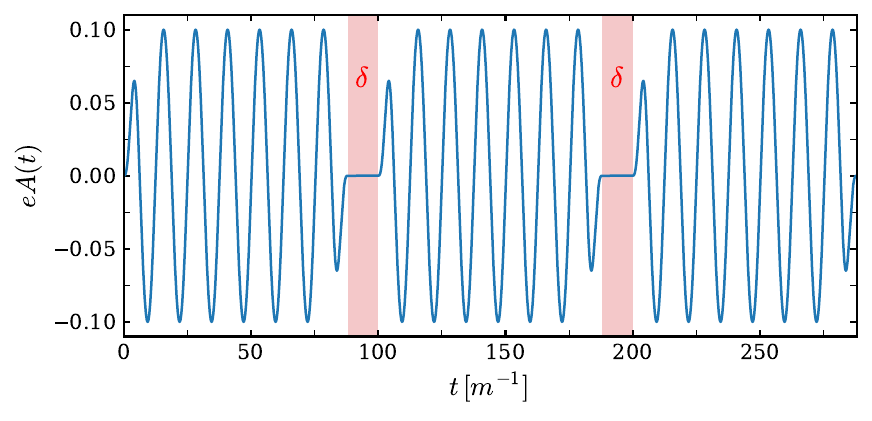}
		\caption{Vector potential \eqref{eq:A(t)_vector_potential}, multiplied by $e$, consisting of three identical pulses ($K=3$) separated by inter-pulse delay of duration $\delta = 12m^{-1}$ (shown in red regions). Each pulse contains $N=6$ carrier cycles with central frequency $\omega = 0.49072\, m$ and peak intensity parameter $\xi = 1$.}
		\label{fig:field}
	\end{figure}

	By construction, $F(t;\tau_p,\tau)$ vanishes outside $[0,\tau_p]$, so each term in the sum in Eq.~\eqref{eq:A(t)_vector_potential} automatically contributes only during its corresponding pulse duration. The total temporal extent of the pulse train is $T = K \, \tau_p + (K-1) \, \delta$. This description produces a train of $K$ smoothly shaped pulses. Fig. \ref{fig:field} shows such a field configuration for $K=3$.
	
	We adopt the same field parameters as used in earlier studies of single- and double-pulse configurations \cite{granz2019electron, mocken2010nonperturbative}, namely
	\begin{equation}\label{eq:field_prameters}
		\xi = 1, \quad \omega = 0.49072m.
	\end{equation}
	These works have shown that electron–positron pair production exhibits pronounced resonances when the ratio of the energy gap to the field frequency becomes an integer. 
	
	The energy gap is given by $2\bar{\epsilon}$, where $\bar{\epsilon}$ denotes the time-averaged quasi-energy of a particle in the external field,
	\begin{equation}\label{eq:quasi-energy}
		\bar{\epsilon} = \frac{1}{T} \int_0^T \sqrt{m^2 + p_x^2 + \bigl(p_y - eA(t)\bigr)^2}\,dt.
	\end{equation}
	For vanishing momenta ($p_x = p_y = 0$) and $\xi = 1$, one obtains $\bar{\epsilon} \approx 1.216007m$. This value is larger than the corresponding field-free energy $\bar{\epsilon}_{\text{free}} = m$ (for $p_x = p_y = 0$), reflecting the dressing of the particle by the external field. 
	The resonance condition is 
	\begin{equation}\label{eq:resonance_condition}
		2\bar\epsilon = n\omega,
	\end{equation} 
	where $n$ is the number of absorbed photons. With the value of $\bar{\epsilon}$ above, the chosen frequency $\omega = 0.49072m$ corresponds to resonant production of particles at rest by absorption of five-photon process ($n = 5$) \cite{mocken2010nonperturbative}.

	\subsection{Dirac equation approach}\label{subsec:dirac_equation}
	Our work uses the time-dependent Dirac equation approach, where the pair production probability in a time-dependent electric field is obtained by solving a coupled system of ordinary differential equations \cite{akal2014electron, otto2015lifting, Grib1972Particle, Bagrov1975Concerning, mocken2010nonperturbative, granz2019electron}. The full derivation is given in \cite{avetissian2002electron, mocken2010nonperturbative}; here, we present only the key steps needed for the numerical implementation. The resulting system is given by
	
	\begin{equation}\label{eq:ODEs}
		\begin{split}
			\dot f(t) &= \kappa(t)f(t) + \nu(t)g(t),\\
			\dot g(t) &= -\nu^*(t)f(t) + \kappa^*(t)g(t),\\
		\end{split}
	\end{equation}
	where
	\begin{equation}\label{eq:ODEs_subparts}
		\begin{split}
			\kappa(t) &= ieA(t)\frac{p_y}{\varepsilon_{\vec{p}}},\\
			\nu(t) &= -ieA(t)\text{e}^{2i\varepsilon_{\vec{p}}t}\left[\frac{(p_x -ip_y)p_y}{\varepsilon_{\vec{p}}(\varepsilon_{\vec{p}} + m)}+i\right].\\
		\end{split}
	\end{equation}
	
	This follows from the time-dependent Dirac equation by adopting the ansatz $\psi_{\vec{p}}(\vec{r},t) = f(t)\,\phi_{\vec{p}}^{(+)}(\vec{r},t) + g(t)\,\phi_{\vec{p}}^{(-)}(\vec{r},t)$, where \(\phi_{\vec{p}}^{(+)}(\vec{r},t) \sim e^{i(\vec{p}\cdot\vec{r} \mp \varepsilon_{\vec{p}} t)}\), with \(\varepsilon_{\vec{p}} = \sqrt{\vec{p}^{\,2} + m^2}\), represent the free Dirac solutions corresponding to momentum \(\vec{p}\) and positive or negative energy states. The validity of this ansatz is based on the conservation of canonical momentum in a spatially homogeneous external field, as ensured by Noether’s theorem. When the vector potential \eqref{eq:A(t)_vector_potential} is on, the canonical momentum reduces to the kinetic momentum $\vec{p}$ of a free particle. As a result, the invariant subspace associated with a fixed momentum $\vec{p}$ can be treated independently using the standard free Dirac states, owing to the rotational symmetry around the field direction, the momentum can be written as $\vec{p} = (p_x,p_y,0)$, where $p_x$ and $p_y$ denote the transverse and longitudinal components, respectively. Furthermore, the presence of a conserved spin-like operator allows an additional reduction of the problem's effective dimensionality \cite{mocken2010nonperturbative, avetissian2002electron}.

	The coefficients $f(t)$ and $g(t)$ introduced in the ansatz represent the occupation amplitudes of the positive- and negative-energy states, respectively. The corresponding system of different equations \eqref{eq:ODEs} is solved with the initial conditions $f(0) = 1$ and $g(0) = 0$. Once the field is switched off, the quantity $f(T)$ gives the occupation amplitude of an electron with momentum $\vec{p}$, positive energy $\varepsilon(\vec{p})$ and certain spin projection. 
	
	Our main interest is the pair production probability for a fixed momentum, which is given by
	\begin{equation}
		W(p_x,p_y) \equiv W(p_x,p_y;T) = 2|f(T)|^2.
	\end{equation}
	The factor of 2 accounts for the two possible spin states. It is useful to note that the created positron has momentum $-\vec{p}$, so that the total momentum of each pair vanishes. 

\section{Numerical results}\label{sec:numerical_results}
	\subsection{Momentum distribution}\label{subsec:momentum_distribution}
	In this section, we present our main numerical results for the pair production probability and the total number of produced pairs in the presence of the field model \eqref{eq:A(t)_vector_potential}. 
	
	\begin{figure*}[tbh]
		\includegraphics[width=\textwidth]{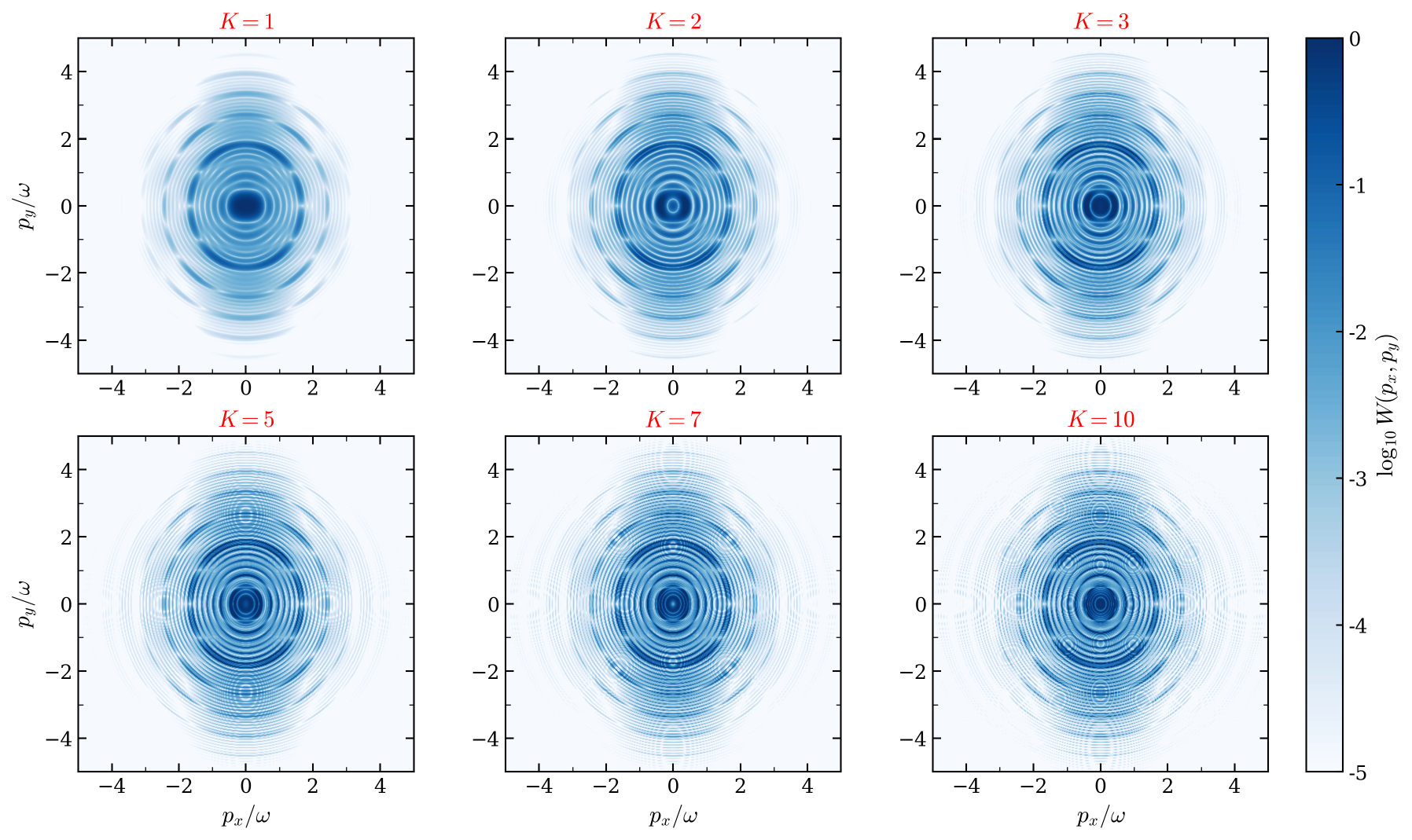}
		\caption{Two dimensional momentum distribution of the produced electrons, shown as $\text{log}_{10}W(p_x,p_y)$, for the multi-pulse field \eqref{eq:A(t)_vector_potential}. The panels are shown for different number of pulses $K$. The field parameters are $\delta = 0, \omega = 0.49072m, \xi =1$ and $N = 6$.}
		\label{fig:2D_W_vs_K}
	\end{figure*}	
	
	Fig. \ref{fig:2D_W_vs_K} shows the characteristic resonance ring structure of the momentum distribution $\text{log}_{10}W(p_x,p_y)$ for a sequence of $K$  pulses. The distribution is symmetric under the $p_{x,y} \rightarrow -p_{x,y}$, leading to an overall reflection symmetry about both the momentum axes. This implies, in particular, that the distribution for electrons and positrons coincide. 
	
	The most prominent feature in all panels is the presence of concentric rings, which constitute the coarse structure of the spectrum. These rings originate from multiphoton absorption processes, where the resonance condition in Eq.~\eqref{eq:resonance_condition} is satisfied for specific momentum values and photon numbers $n$. In the present case, the central region around $p_x = p_y = 0$ is governed by a $5\omega$ resonance, while the outer rings correspond to higher-order channels involving total energy absorption of $6\omega$, $7\omega$, $8\omega$, and so on. Each integer $n$ therefore defines a distinct energy gap, which in turn fixes the radius of the corresponding ring in momentum space. Consequently, the positions of these rings remain unchanged as $K$ is varied. Note that these ring-shaped structure are not perfectly circular but slightly elongated along the longitudinal direction. This deviation from circular symmetry originates from the way the longitudinal momentum enters the quasienergy through the combination $p_y - eA$. Consequently, larger values of $p_y$ can be partially compensated by the external field, effectively lowering the energy cost for pair production in that direction. This leads to an enhanced probability for producing particles with higher longitudinal momentum, causing the rings to stretch and appear elongated.
	
	For multiple pulses ($K>1$), the distribution undergoes a qualitative change and develops increasingly pronounced modulations, becoming more structured and less smooth. In particular, the initially broad rings observed for $K=1$ progressively narrow with increasing $K$, reflecting the improved energy resolution associated with a longer effective interaction time. At the same time, fine ring substructures emerge and become superimposed on the coarse rings. These finer features grow in number and sharpness as $K$ increases, indicating the buildup of temporal coherence over the extended duration of the pulse sequence. As a result, each multiphoton resonance ring evolves from a broad feature into a set of narrow, finely modulated structures in momentum space.
	\begin{figure*}[tbh]
		\includegraphics[width=\textwidth]{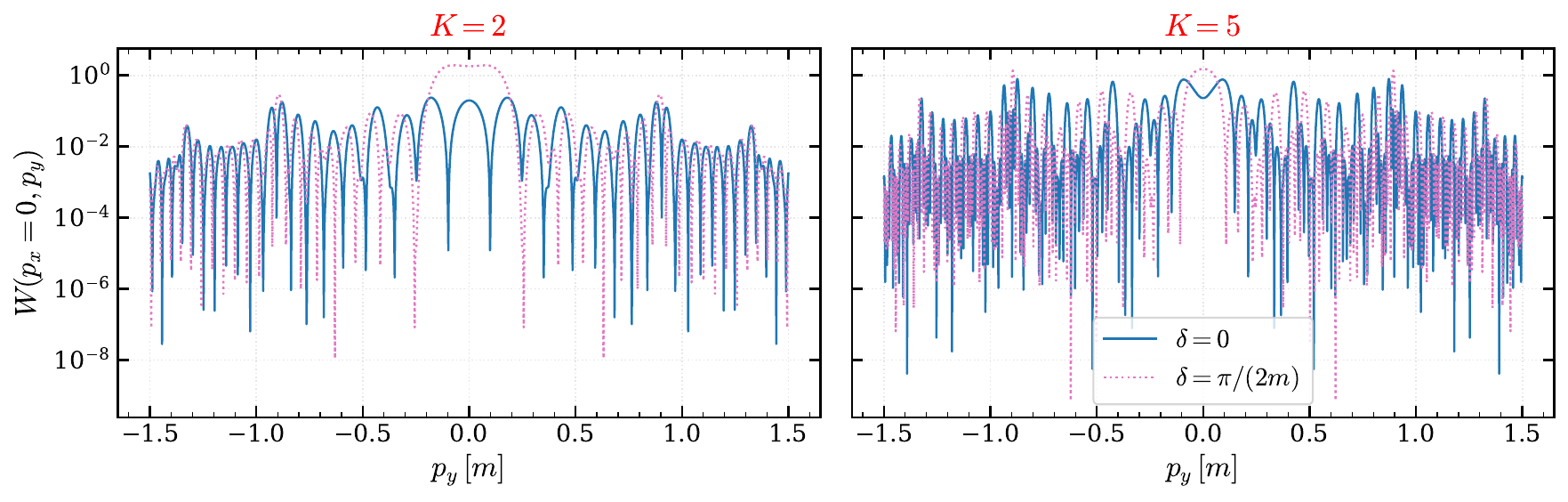}
		\caption{Longitudinal momentum distribution $W(p_x = 0, p_y)$ of the produced electrons in presence of the multi-pulse field \eqref{eq:A(t)_vector_potential} for different number of pulses $K$ and inter-pulse delay $\delta$. The other field parameters are  $\omega = 0.49072m, \xi =1$ and $N = 6$.}
		\label{fig:1D_W_vs_py}
	\end{figure*}
	
	To illustrate the effect of the inter-pulse delay $\delta$ on the momentum distribution, Fig.~\ref{fig:1D_W_vs_py} presents the longitudinal momentum spectrum $W(p_x = 0, p_y)$ of the produced electrons for different values of the pulse number $K$ and delay $\delta$. In all the cases, the spectrum exhibits a strongly oscillatory pattern consisting of alternating peaks and nodes. As $K$ increases, these oscillations become significantly denser and sharper, indicating the longer effective interaction time. For the finite delay $\delta = \pi/(2m)$, we observe the redistribution of the spectral weights, shifting in the peak positions, and stronger suppression at certain momenta. 
	
	\subsection{Total number of pairs}\label{subsec:number_density}
	\begin{figure}[tbh]
		\includegraphics[width=0.5\textwidth]{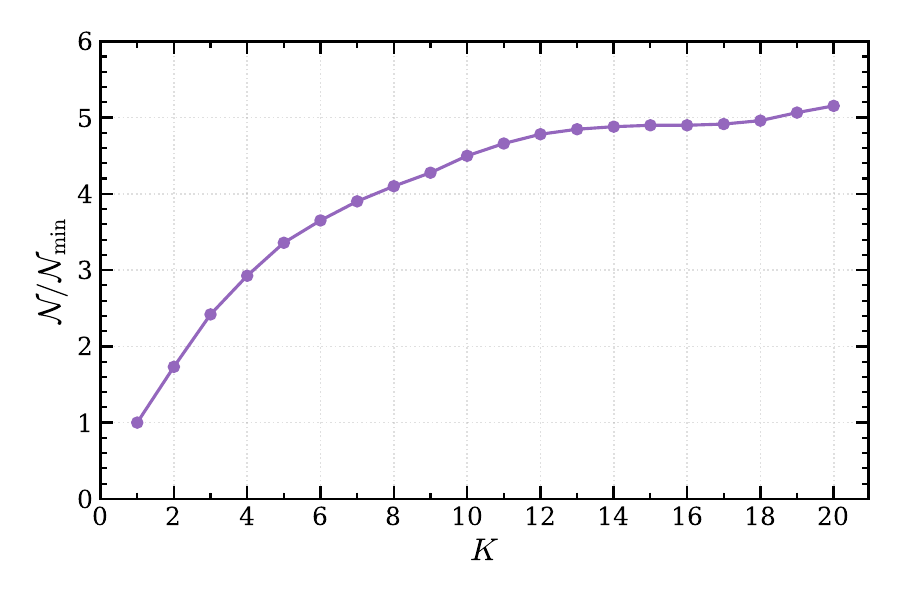}
		\caption{Total number of produced pairs per Compton volume ($1/m^{3}$), normalized by dividing through the minimum of $\mathcal{N}$, as a function of number of pulses $K$ for the multi-pulse field configuration \eqref{eq:A(t)_vector_potential}. The minimum number of produced pairs is $\mathcal{N}_\text{min} = 2.64\times10^{-4}$. The field parameters are $\delta = 0, \omega = 0.49072m, \xi = 1$ and $N  = 6$.}
		\label{fig:total_number_vs_K}
	\end{figure}

	The total number of produced pairs can be obtained by integrating the momentum dependent production probability $W(p_x,p_y)$ over the entire momentum space. Since the field model in Eq. \eqref{eq:A(t)_vector_potential} is directed along the $\hat{y}$ axis, the system exhibits cylindrical symmetry. Consequently, the numerical evaluation of the integral reduces to \cite{mocken2010nonperturbative, brass2025relative}
	\begin{equation}\label{eq:total_number}
		\mathcal{N} \approx \frac{1}{4\pi^2} \int^{p_y^\text{(max)}}_{-p_y^\text{(max)}} \int^{p_x^\text{(max)}}_0 W(p_x,p_y)p_x dp_x dp_y
	\end{equation}
	where ${p_x^\text{(max)}}$ and ${p_y^\text{(max)}}$ are chosen sufficiently large to capture the region where pair production is significant. 
	
	Fig. \ref{fig:total_number_vs_K} shows the total number of pairs produced per Compton volume $(1/m^3)$ as a function of the pulse number $K$. We observe a clear trend of rapid initial growth followed by gradual saturation. For small $K$, the number density increases steeply, which indicates that adding more pulses significantly enhances pair production due to increased interaction time. As $K$ continues to grow, the curve shows slow increase trend, suggesting that the system is entering a regime where additional pulses contribute less effectively. We believe that this saturation-like behavior is a consequence of phase averaging and partial destructive interference between widely separated pulses, which limits further coherent enhancement. The slight increase observed at larger $K$ indicates that coherence effects are still present but become progressively weaker, leading to diminished pair production efficiency.
	
	\subsection{Multi-slit interference}
	The multi-pulse oscillatory time-dependent electric field defined in Subsec. \ref{subsec:field_profile} can be interpreted as a multi-slit interferometer in time domain, where each individual pulse acts as a coherent source of electron-positron pairs. This picture follows the framework developed in \cite{akkermans2012ramsey, li2014enhanced, li2014multiple}, where temporally separated pulses generate amplitudes that interfere analogously to waves passing through multiple spatial slits. 
	
	Within the quantum kinetic or scattering formulation \cite{akkermans2012ramsey}, the total production amplitude can be expressed as a coherent sum over contributions from each pulse. For a sequence of $K$ identical pulses, this can be written as
	\begin{equation}\label{eq:amplitude_sum}
		\mathcal{A}_K \sim  \mathcal{A}_1 \sum_{k=0}^{K-1}e^{i\phi_k},
	\end{equation}
	where $\mathcal{A}_1$ denotes the amplitude of a single pulse and $\phi_k$ represents the dynamical phase accumulated between successive pulses. For equally spaced pulses, the phase difference between consecutive pulses becomes approximately constant, $\phi_k \sim k\Delta\varphi$, this leads to
	\begin{equation}\label{eq:interference_formula}
		\begin{split}
			W_K \sim |\mathcal{A}_K|^2 &= |\mathcal{A}_1|^2\left|\sum_{k=0}^{K-1}e^{ik\Delta\varphi}\right|^2\\
			 &= W_1\frac{\text{sin}^2(K\Delta\varphi/2)}{\text{sin}^2(\Delta\varphi/2)}.
		\end{split}
	\end{equation}
	This is the standard multi-slit interference formula, identical in form to optical diffraction from $K$ slits. 
	
	\begin{figure}[tbh]
		\includegraphics[width=0.5\textwidth]{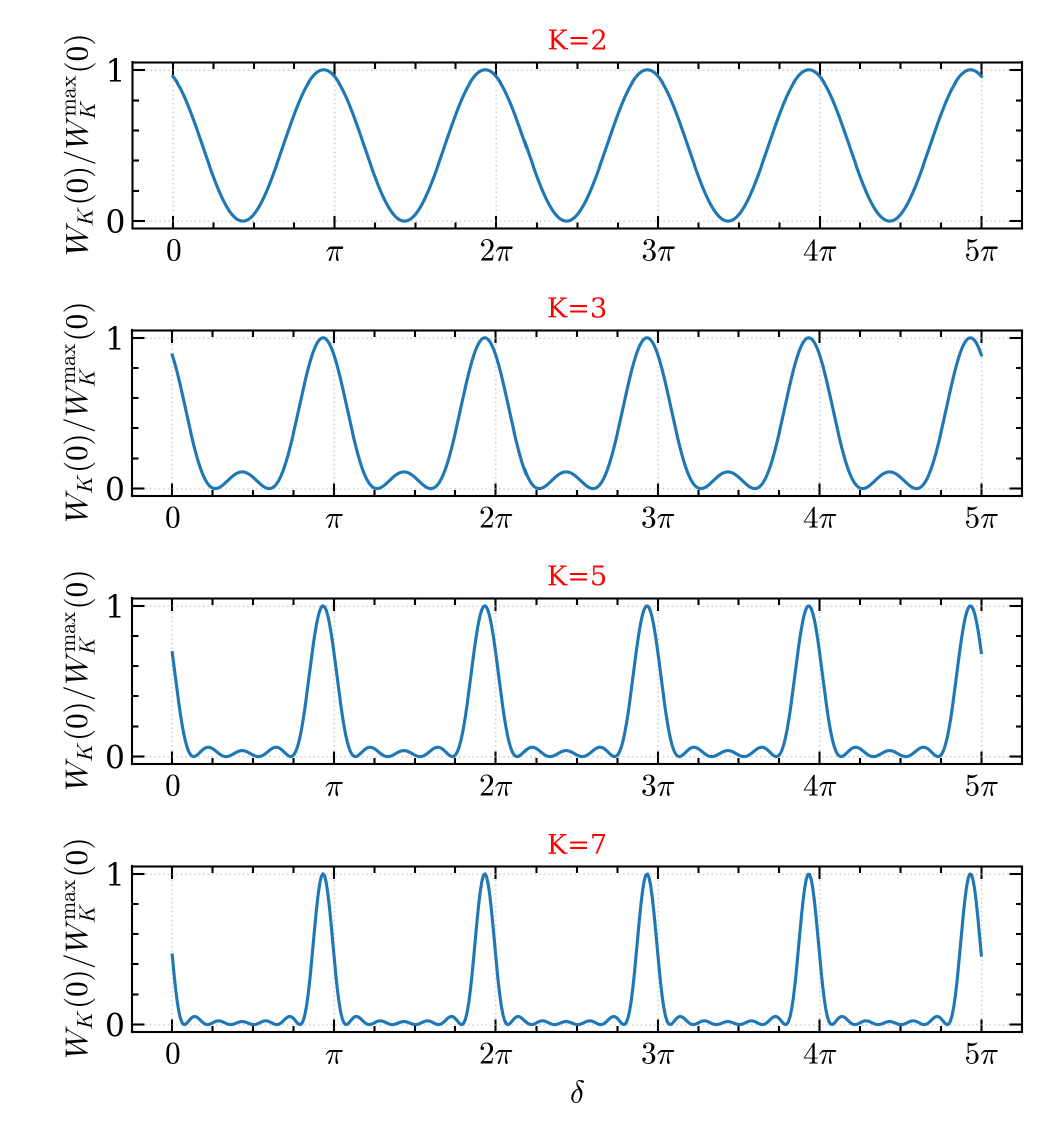}
		\caption{Pair production probability at zero momentum, $W_K(0)$, normalized to its 
			maximum value, i.e., $W_K(0)/W_K^{\max}(0)$, as a function of the inter-pulse 
			delay $\delta$, for a multi-pulse field \eqref{eq:A(t)_vector_potential} with pulse number $K$. The field parameters 
			are $\omega = 0.5m$, $\xi = 0.1$ and $N = 6$.}
		\label{fig:W_vs_delta} 
	\end{figure}
	
	In order for this quantum interference picture of coherence of amplitudes to remain valid, it is essential that the pair production probability remains small i.e., $W\ll 1$ \cite{akkermans2012ramsey}. To ensure this condition, we choose $\xi = 0.1$. For a fixed momentum $\vec{p}$, the pair production probability depending on the inter-pulse delay $\delta$ via the standard Fabry-Perot form is shown in Fig. \ref{fig:W_vs_delta}. For $K=2$, the interference reduces to a simple two-path interference giving a smooth sinusoidal oscillations in $\delta$. On increasing $K$, the interference factor develops sharper principal maxima, and the pattern becomes increasingly structured representing the Ramsey-type interference fringes in the time domain, with $\delta$ playing the role of the interferometric control parameter.  At specific values of $\delta$, constructive interference leads to strong enhancement of the pair production probability, while for other values destructive interference suppresses it. 
	
	At the principal maxima, where all phases align constructively, the production probability exhibits an approximately quadratic scaling with the number of pulses, $W\propto K^2$. This behavior follows from the interference relation in Eq.~\eqref{eq:interference_formula} and is supported by the numerical results presented in Table~\ref{tab:quadratic_scaling}.
	This clearly demonstrates the coherent nature of the process and highlights the role of temporal interference in enhancing pair production.
	
	\begin{table}[tbh]
		\caption{Quadratic scaling of pair production probability $(W \propto K^2)$ with pulse number $K$. The table corresponds to the Fig. \ref{fig:W_vs_delta}.}
		\label{tab:quadratic_scaling}
		\begin{ruledtabular}
			\begin{tabular}{ccc}
				$K$ & $W^\text{max}_K(0)$ & $W^\text{max}_K(0)/W_1(0)$ \\
				\hline
				1 & $1.265\times10^{-7}$ & $1$ $(=1^2)$ \\
				2 & $5.061\times10^{-7}$ & $4$ $(=2^2)$ \\
				3 & $1.139\times10^{-6}$ & $9$ $(=3^2)$ \\
				5 & $3.162\times10^{-6}$ & $24.99$ $(\approx5^2)$ \\
				7 & $6.195\times10^{-6}$ & $48.97$ $(\approx7^2)$ \\
			\end{tabular}
		\end{ruledtabular}
	\end{table}
	
\section{Conclusion}\label{sec:conclusion}

	In this work, we investigated electron–positron pair production in a time-dependent oscillating electric field with a multi-pulse structure by numerically solving the time-dependent Dirac equation. The resulting momentum spectra exhibit characteristic multiphoton ring structures, where each ring corresponds to a resonance associated with the absorption of an integer number of photons. As the number of pulses increases, these rings become progressively thinner, indicating improved energy resolution due to the longer effective interaction time. At the same time, additional finer rings appear within the coarse structure, reflecting the buildup of temporal coherence and interference effects.	
	
	In addition, the inter-pulse delay plays an important role in shaping the momentum distribution, leading to shifts in peak positions, redistribution of spectral weight, and suppression at specific momenta. This demonstrates that the temporal spacing between pulses provides an effective control parameter for tailoring the pair production process. We further showed that the total number of produced pairs increases with the number of pulses. For small pulse numbers, the relative enhancement is significant, while for larger values a slower growth is observed, suggesting the onset of saturation due to phase averaging and partial destructive interference. 
	
	Finally, the multi-pulse configuration can be interpreted as a time-domain Ramsey interferometer, where each pulse acts as a coherent source of pairs. We showed the Ramsey-type fringes in the production probability as a function of inter-pulse delay, along with an approximate quadratic scaling with the number of pulses at constructive interference points. These results highlight the role of coherent temporal interference as a mechanism for controlling and enhancing vacuum pair production in strong-field QED.  

\bibliographystyle{apsrev4-2}  
\bibliography{biblio}
\end{document}